\documentclass[11pt]{article}
\usepackage{xspace,amssymb,amsmath,helvet,graphicx}
\begin{document}
\newcommand{\dee}{\,\mbox{d}}
\newcommand{\naive}{na\"{\i}ve }
\newcommand{\eg}{e.g.\xspace}
\newcommand{\ie}{i.e.\xspace}
\newcommand{\pdf}{pdf.\xspace}
\newcommand{\etc}{etc.\@\xspace}
\newcommand{\PhD}{Ph.D.\xspace}
\newcommand{\MSc}{M.Sc.\xspace}
\newcommand{\BA}{B.A.\xspace}
\newcommand{\MA}{M.A.\xspace}
\newcommand{\role}{r\^{o}le}
\newcommand{\signoff}{\hspace*{\fill} Rose Baker \today}
\newenvironment{entry}[1]%
{\begin{list}{}{\renewcommand{\makelabel}[1]{\textsf{##1:}\hfil}%
\settowidth{\labelwidth}{\textsf{#1:}}%
\setlength{\leftmargin}{\labelwidth}
\addtolength{\leftmargin}{\labelsep}
\setlength{\itemindent}{0pt}
}}%
{\end{list}}
\title{A new generalization of the beta distribution}
\author{Rose Baker\\School of Business\\University of Salford, UK}
\maketitle
\begin{abstract}
The beta distribution is the best-known distribution for modelling doubly-bounded data, \eg percentage data or probabilities.
A new generalization of the beta distribution is proposed, which uses a cubic transformation of the beta random variable.
The new distribution is label-invariant like the beta distribution and has rational expressions for the moments. This facilitates its use in mean regression.
The properties are discussed, and two examples of fitting to data are given. A modification is also explored in which the Jacobian of the transformation is omitted.
This gives rise to messier expressions for the moments but better modal behaviour. 
In addition, the Jacobian alone gives rise to a general quadratic distribution that is of interest.
The new distributions allow good fitting of unimodal data that fit poorly to the beta distribution, and could also be useful as prior distributions.
\end{abstract}
\section*{Keywords}
Beta distribution, cubic transformation, doubly-bounded data, Jacobian, label invariance, regression
\section{Introduction}
Doubly-bounded data occur in many application areas, as for example percentage data, and also when the random variable is a probability.
In this case, the probabilities are not often directly measured, and the data are binary, with events such as death that either occur or do not.
However, a distribution of probability may then be used as a prior distribution, in a Bayes or Empirical Bayes analysis.

There are relatively few distributions available for modelling doubly-bounded data. The 2-parameter beta distribution defined on $[0,1]$ has pride of place, and many attempts have been made to
generalize it to allow more flexible behaviour. These include the generalized beta distribution
reviewed by Pham-Gia and Duong (1989). Other generalizations are considered by Nadarajah and Kotz (2006).
There are also two very simple generalizations. One is a mixture with a uniform distribution, to allow greater variance. Its use in regression is described in Bayes, Baz\'{a}n and Garcia (2012).
Another is zero inflation, a mixture with a delta-function at zero (Stewart, 2013).

Attempts have also been made to replace the beta distribution with a more flexible distribution. The best-known alternative is probably the Kumaraswamy distribution ({\em ibid}, 1980).
Some other replacement distributions are described in the book `Beyond Beta' by Kotz and van Dorp (2004), and include 2-sided power distributions, generalized trapezoidal distributions,
the Topp and Leone distribution, and Johnson's $S_B$ distribution (for which see Johnson, Kotz and Balakrishnan, 1995). There is also the log-Lindley distribution (G\'{o}mez-D\'{e}niz {\em et al}, 2014).

Many of these distributions are difficult to use, with likelihoods and moments only expressible in terms of hypergeometric functions, or with cusps (the 2-sided power and trapezoidal distributions).
No useful distribution can be completely simple, and even the normal distribution requires a special function, the error function, to compute its distribution function.
However, there are several requirements for a practically useful distribution to generalize the beta distribution. 

One is that the moments should be simple to compute, or at least the mean.
This is because we often wish to regress the mean on a covariate, \eg mean percentage body fat can be regressed on the Quetelet index (body mass index or BMI).
To do this using a likelihood-based method, we must compute the mean $\mu$ for an observation, as a function of the covariates, and in order to compute the log-likelihood, we must then be able to compute the model parameters.
For example, with the beta distribution itself with parameters $\alpha, \beta$, the mean is $\mu=\alpha/(\alpha+\beta)$, and Mielke (1975) suggests reparameterising to use $\mu$ and $s=\alpha+\beta$.
Then we  compute $\alpha=s\mu$ and $\beta=s(1-\mu)$ and can compute the log-likelihood. Without a simple formula for the mean in terms of the model parameters,
such a regression would be difficult.

Another requirement is label-invariance. With a random variable $X$, and standardising the interval to be be $[0,1]$, we can look at the distribution of $Y=1-X$, where the ends of the scale have been flipped.
We have interchanged the labels, for example `success' and `failure', or `no treatment effect' and `100\% treatment effect'. Label-invariance means that $Y$ follows a distribution from the same family as $X$.
Thus if $X \sim Beta(\alpha,\beta)$, then $Y \sim Beta(\beta,\alpha)$. The fitted model will be the same with the same likelihood value and the same fitted parameter values
whichever choice is made.

Some distributions, such as the Kumaraswamy distribution, are not label-invariant. However, label invariance in general is a common requirement \eg in medicine. (\eg Senn, 1996). For example, suppose we wish to model the distribution of a disease activity index measured on a scale from zero to unity.
Without a label-invariant model, we would get a different distribution if we considered the corresponding
health index $Y=1-X$. Which model should we believe? 

We must also require that the pdf and distribution function can be easily computed on most platforms, \ie they do not require special functions that may not be available;
this is not a crucial requirement, because if a distribution proves useful, the necessary special functions will soon be produced.
The distribution function is needed for computing the likelihood when data are censored, an extreme case being the application of a doubly-bounded distribution to
fitting grouped data. Random numbers are also needed, for example in Markov-chain Monte Carlo methods, and their generation should preferably be straightforward.

The beta distribution itself has the virtues of having a simple expression for the mean and of label-invariance. To compute the likelihood we require the beta function, and the distribution function requires the incomplete beta function.
This is also needed for the t-distribution, so is commonly available. Random number generation is not particularly simple, but there exist efficient methods for doing this.

It was desired to construct a practically useful generalized beta distribution, as there does not currently appear to be a distribution for doubly-bounded data
that has relatively simple expressions for the moments, is free of cusps, and is label-invariant.
The log-Lindley distribution, for example, has simple expressions for the moments but is not label-invariant.

Bearing the above considerations in mind, we took the random variable $X$ as $X=\sum_{j=1}^m c_j P^j$, where $P \sim Beta(\alpha,\beta)$, and the $c_j$ are chosen
so that $\dee x/\dee p > 0$, \ie we have a monotonic transformation of the beta random variable.
This has been explored with $m=2$ and $m=3$. The case $m=1$ of course simply gives the beta distribution.

These new distributions generalize the beta distribution and allow more flexible behaviour, \eg the skewness for a given mean can change substantially.
They are also label-invariant, as we can write $Y=1-X=\sum_{j=1}^m c_j^{\prime}(1-P)^j$, where the $c_j^{\prime}$ are linear functions of the $c_j$.
The mean can also be computed as a rational expression, thus facilitating its regression on covariates.

The difficult task is restricting the $c_j$ to require that $\dee x/\dee p > 0$. We shall see that this is straightforward for $m=2$ (quadratic or Q-beta distribution) and less so for $m=3$
(cubic or C-beta distribution).
We have not yet gone beyond $m=3$, where we already have two extra parameters, which gives plenty of flexibility.

An unexpected problem with the $m=3$ distribution led to the creation of `Jacobian-less' distributions, and these are regarded as the most useful distributions resulting from this approach.

The next section briefly discusses transformations of  distributions, then the following sections discuss the detailed properties of the new distributions, after which two examples of their use are given.
Finally, the quadratic distributions arising when the transformation is applied to the uniform distribution are described in more detail in appendices.
\section{Transformation of pdfs}
Consider a distribution with pdf $f_p(p)$ defined on $[0,1]$ and a transformation to $x=x(p)$ with an inverse transformation $p=p(x)$. Here $f_p$ is the pdf of the beta distribution with parameters $\alpha, \beta$, and $x(p)=ap+bp^2+cp^3$.
Denote the respective random variables as $P$ and $X$ respectively.

We require that the Jacobian $J(x)=\dee x/\dee p > 0$, so that the transformation is monotonic and one-to-one. Since $\text{Prob}(X < x)=\text{Prob}(P < p)$ we have that
$f_x(x(p))=f_p(p) \dee p/\dee x=f_p(p)/(\dee x/\dee p)$.

If the Jacobian becomes very small over some interval of $X$, $f_x(x)$ will become large, and the distribution might be multimodal. In view of this, a `Jacobian-less'
distribution was constructed with pdf $g$, for which $g_x(x(p))=Cf_p(p)$ for some unknown constant $C$. This distribution clearly has the same modal structure as $f_p(p)$, because
$\dee g_x(x)/\dee x=C(\dee f_p(p)/\dee p)/(\dee x/\dee p)$. Since $\dee x/\dee p > 0$, the mode of $g_x(x)$ occurs at $x_m=x(p_m)$, where $p_m$ is the mode of $f_p$.
If $C$ could not be easily determined, this type of pdf would be of little interest. The `parent' distribution of $P$ which is transformed to yield $g_x(x)$
is $Cf_p(p)\dee x/\dee p$, which here is simply $Cf_p(p)(a+2bp+3cp^2)$. This is a mixture of beta-distributions, although some of the weights in the mixture may be negative.
It is therefore straightforward to evaluate $C$ by requiring that the pdf integrates to unity.

Discarding the Jacobian ensures a unimodal distribution if $f_p(p)$ is unimodal. However, it unavoidably makes results for distribution functions and moments more complicated,
and the motto when considering discarding Jacobians should be `if it's not broken, don't fix it'.
\section{Computing issues and notation}
To solve cubics, and even quadratics, use of the Newton-Raphson iteration (\eg Press {\em et al}, 2007) is often recommended here in preference to analytic solutions. This is very quick and easy to program.

If it happens that the variable $x$ strays outside $[0,1]$ this probably would not cause a problem, because it would wander back into $[0,1]$ again before convergence, but it is faster and safer to set
$x \rightarrow \text{min}(\text{max}(x,0),1)$ after each step. It is safer because $\dee x/\dee p$ is not guaranteed to be positive for $p < 0$ or $p > 1$ and so the iteration might oscillate or diverge.

Computation was done using purpose-written fortran programs and the NAG library.

To define some notation, let $P$ be a r.v. that follows the beta distribution with parameters $\alpha, \beta$, so that the pdf $f(p)$ is:
\[f_p(p)=p^{\alpha-1}(1-p)^{\beta-1}/B(\alpha,\beta),\]
where $B$ denotes the beta function. We sometimes write for brevity $\eta=\alpha+\beta$.
Distributions are taken as having support in $[0,1]$, but it is trivial to change the interval to an arbitrary interval by adding two more parameters.

Note that quantiles must always be found from the distribution function using Newton-Raphson iteration; they are not discussed further.
It is also straightforward to compute inverse moments such as $\text{E}(1/X)$ or $\text{E}\{(1-X)/X\}$; these are also not discussed further.
Bivariate distributions could be constructed, but currently the best procedure would be to use a copula.

\section{Properties: the Q-beta (quadratic) distribution}
It seems that the $m=3$ (cubic) distribution is a lot more flexible than the $m=2$ distribution, and the Jacobian-less distribution still better, but we consider the simpler cases first.
\subsection{Pdf}
Define $X=2\gamma P+(1-2\gamma)P^2$ where $0 < \gamma < 1$, where $a=2\gamma, b=1-2\gamma, c=0$. We say that $X$ follows the Q-beta (quadratic-beta) distribution, \ie
$X \sim QB(\alpha,\beta,\gamma)$.
First, since $x(p)=2\gamma p+(1-2\gamma)p^2$, we have that $x(0)=0, x(1)=1$ and $\dee x/\dee p=2\gamma+2(1-2\gamma)p$. This quickly leads to the requirement $0 \le \gamma \le 1$,
and when $\gamma=1/2$ we regain the beta distribution.
The pdf $f_x(x)=f(p)\dee p/\dee x$. Solving the quadratic $2\gamma p+(1-2\gamma)p^2-x=0$ gives
\[p(x)=\frac{-\gamma+\Delta(x)}{1-2\gamma}=\frac{x}{\gamma+\Delta(x)}\] where $\Delta(x)=(\gamma^2+(1-2\gamma)x)^{1/2}$, and $\dee x/\dee p=2\Delta(x)$.
Hence the pdf $f_x(x)$ is
\[f_x(x)= (\gamma+\Delta(x))^{2-\alpha-\beta}x^{\alpha-1}(\gamma+\Delta(x)-x)^{\beta-1}/2\Delta(x) B(\alpha,\beta).\]
The distribution, like the beta distribution, is label-invariant, so that if $X \sim QB(\alpha,\beta,\gamma), 1-X \sim QB(\beta,\alpha,1-\gamma)$.
We have $1-x=2(1-\gamma)(1-p)+(1-2(1-\gamma))(1-p)^2$, and so $1-p=\frac{1-x}{1-\gamma+\Delta(x)}$ and we have the alternative form
\[f_x(x)=\frac{x^{\alpha-1}(1-x)^{\beta-1}}{2\Delta(x) B(\alpha,\beta)(\gamma+\Delta(x))^{\alpha-1}(1-\gamma+\Delta(x))^{\beta-1}}.\]

In practice to compute the pdf a less analytic approach, which works well also for the cubic distribution considered in the next section, can be used.
We find $p(x)$ by solving $2\gamma p+(1-2\gamma) p^2-x=0$ either by solving the quadratic or by Newton-Raphson iteration from $p=x$ and the pdf is then $f(p)/(2(\gamma+(1-2\gamma)p)$.
\subsection{Distribution function}
The distribution function is also simply related to the distribution function of the beta distribution: using $f_x(x)\dee x=f(p)\dee p$ yields $F(x)=I(\alpha,\beta;p(x))=I(\alpha,\beta;x/(\gamma+\Delta(x)))$, where $I(\alpha,\beta;p(x))$
denotes the incomplete beta function ratio.
\subsection{Random numbers}
Random numbers are generated by piggy-backing off the beta distribution;  generate $P \sim Be(\alpha,\beta)$ and then $X=2\gamma P+(1-2\gamma)P^2$.
\subsection{Moments}
The moments are also simple to calculate, although messy. We have that
\[\text{E}(X^n)=\int_0^1 f_x(x) x^n \dee x=\int_0^1 (2\gamma p+(1-2\gamma))p^2)^n f(p) \dee p,\]
from which the moments can be read off using 
\[\int_0^1 p^m f(p)\dee p=(\alpha)_m/(\alpha+\beta)_m,\]
where $(a)_m$ denotes the Pochhammer symbol, the ascending factorial, so that $(a)_m=a(a+1)\cdots(a+m-1)$.
Finally,
\[\text{E}(X^n)=\sum_{j=0}^n {n \choose j}(1-2\gamma)^j(2\gamma)^{n-j}\frac{(\alpha)_{n+j}}{(\alpha+\beta)_{n+j}}.\]
Specifically,
\[\text{E}(X)=\frac{\alpha(2\gamma\beta+\alpha+1)}{(\alpha+\beta)(\alpha+\beta+1)}.\]
\[\text{var}(X)=4\gamma^2(\alpha)_2/(\alpha+\beta)_2+4\gamma(1-2\gamma)(\alpha)_3/(\alpha+\beta)_3+(1-2\gamma)^2(\alpha)_4/(\alpha+\beta)_4-\text{E}(X)^2.\]
\subsection{Mode}
The mode is best found from the pdf expressed in terms of $p(x)$ as $f_x(x)=p^{\alpha-1}(1-p)^{\beta-1}/2(\gamma+(1-2\gamma)p)B(\alpha,\beta)$.
Taking logarithms and writing $\dee \ln f/\dee x=(\dee \ln f/\dee p) (\dee p/\dee x)$ the mode is at $x_m=2\gamma p_m+(1-2\gamma)p_m^2$, where
\[\frac{\alpha-1}{p_m}-\frac{\beta-1}{1-p_m}-\frac{1-2\gamma}{\gamma+(1-2\gamma)p_m}=0\]
under the same conditions as for the mode of the beta distribution.

One can find $p_m$ and hence $x_m$ by solving this equation by Newton-Raphson iteration  or by solving the corresponding quadratic
\[(1-2\gamma)p^2+\{(\alpha-2)(1-2\gamma)-(\alpha+\beta-2)\gamma)p+(\alpha-1)(1-\gamma)=0.\]
For $\alpha >1, \beta > 1$ the distribution is unimodal and in general has exactly the same modality as the beta distribution.

\section{Properties: the C-beta (cubic) distribution}
\subsection{Parameters}
We can add two parameters and create the variable $x(p)=ap+bp^2+cp^3$.
Since $x(1)=1$, $a+b+c=1$, and we focus on $c$ and $a$, which will yield parameters $\gamma$ and $\delta$. We require $\dee x/\dee p \equiv J(p)>0$, so $a+2bp+3cp^2 > 0$,
whence $a+2b+3c > 0$ or $a < c+2$.  Given $a > 0, a < c+2$, then
$J(p) > 0$ at $p=0$ and $p=1$ and $J(p)$ could only become negative if the equation for zero slope, $a+2bp+3cp^2=0$,
has at least one root in $[0,1]$. A sufficient condition for this not to happen is that the determinant $\Delta^2(x)=b^2-3ac < 0$.
Substituting for $b=1-a-c$ this yields limits for $a$ of $1+(1/2)c\pm\sqrt{3c(4-c)}/2$.
Note by the way that when $c=4$, $a=3$, and that we must have $-2 \le c \le 4$.

This last can be seen more convincingly from $J(1/2) > 0$, \ie $a+b+(3/4)c > 0$ or $c < 4$.
From $J(2/3)=a+(4/3)b+(4/3)c > 0$, \ie $3a+4b+4c > 0$, we have $a < 4$.

The requirement that $\Delta^2 < 0$ is not a necessary condition for $J(p) > 0$ in $[0,1]$, as real roots may exist but be outside the range $[0,1]$.
When $c \ge 1$, the lower limit is as previously given, and when $ c \le 1$, the upper limit for $a$ is $c+2$ and the lower limit zero.
This can be seen as follows: write $J(p)=a(1-p)^2+2(a+b)p(1-p)+(a+2b+3c)p^2$, \ie an expansion in Bernstein polynomials. Clearly all coefficients are positive when $a < c+2$ (so that $a+2b+3c > 0$) if $a+b > 0$, \ie $c < 1$.
Hence for $c < 1$ the upper limit for $a$ such that $J(p) > 0$ in $[0,1]$ is $c+2$ and the lower limit is zero.

To summarise:
\begin{enumerate}
\item $-2 \le c \le 4$;
\item if $c \le 1$, the range of $a$ is $(0, c+2)$;
\item if $c \ge 1$, the range of $a$ is $1+c/2-\sqrt{3c(4-c)}/2, 1+c/2+\sqrt{3c(4-c)}/2)$.
\end{enumerate}
We now consider how $a$ and $c$ transform when $X \rightarrow 1-X$. Writing $1-X=a^\prime (1-P)+b^\prime (1-P)^2+c^\prime (1-P)^3$
we have that $a^\prime=a+2b+3c$, so that $a^\prime=2+c-a$, and $c^\prime=c$. Hence $c$ is invariant under the label transformation, and we take the first parameter $\delta=(c+2)/6$,
so that $0 \le \delta \le 1$. Next, for $c < 1$ (or $\delta < 1/2$), define $\gamma=a/(c+2)$, and for $c > 1$ define $\gamma=\frac{a-(1+c/2)}{\sqrt{3c(4-c)}}+1/2$,
so that $0 \le \gamma \le 1$ always. Under the label transformation, $\delta$ stays the same, while $\gamma \rightarrow 1-\gamma$.
Thus $\text{C-beta}(\alpha,\beta,\gamma,\delta)\rightarrow \text{C-beta}(\beta,\alpha, 1-\gamma, \delta)$.
Distributions with $\gamma=1/2$ and $\alpha=\beta$ are therefore symmetric.

This parameterisation allows $\gamma, \delta$ to each vary in $[0,1]$ and exhibits the label symmetry, but has the drawback
that the model with $c=0$ will not have a zero value of $\delta$; it occurs at $\delta=1/3$.
To convert $\gamma,\delta$ to $a, b, c$, we write $c=6\delta-2$, then if $\delta < 1/2$, $a=(c+2)\gamma$, otherwise $a=(\gamma-1/2)\sqrt{3c(4-c)}+1+c/2$,
and finally $b=1-a-c$.  When $\delta=1/3$ so that $c=0$ we have $a=2\gamma$. Hence $\gamma$ has the same meaning as for the $\text{Q-beta}$ distribution,
which is now seen to be the $\text{C-beta}$ distribution with $\delta=1/3$.

A sensible fitting sequence to ensure convergence would be:
\begin{enumerate}
\item fit the beta distribution in the usual way;
\item fit the Q-beta distribution starting from the fitted $\alpha$ and $\beta$ values, with $\gamma=1/2$ (so that $a=1$), $\delta=1/3$ (so that $c=0$).
\item in case of difficulty, float $\gamma$ and then $\delta$, or vice versa.
\end{enumerate}

\subsection{Pdf}
To obtain $p$ from $x$ we solve the cubic $cp^3+bc^2+ap-x=0$ to obtain $p(x)$. This could be done analytically, \eg using Vieta's method,
but for some parameter values this method is numerically unstable, and a Newton-Raphson iteration starting from $p=x$ is fast and always converges quickly
with no numerical problems. 

The pdf can then be computed as 
\begin{equation}f_x(x)=\frac{p(x)^{\alpha-1}(1-p(x))^{\beta-1}}{B(\alpha,\beta)(3cp(x)^2+2bp(x)+a)}.\label{eq:cubf}\end{equation}
\subsection{Distribution function}
We have again $F(x)=I(\alpha,\beta;p(x))$.
\subsection{Moments}
The moments are found using the identity
\[\text{E}(X^n)=\int_0^1f(p)(a+bp+cp^2)^n \dee p,\]
where $a, b, c$ are found from $\gamma, \delta$.
The mean is
\[\text{E}(X)\equiv \mu=\frac{\alpha}{\alpha+\beta}\{a+\frac{\alpha+1}{\alpha+\beta+1}\{b+\frac{\alpha+2}{\alpha+\beta+2}c\}\},\]
which has been arranged for fast computation.

When regressing the mean on covariates, one can proceed as follows. 
\begin{enumerate}
\item take as parameters $\mu, \eta=\alpha+\beta, \gamma$ and $\delta$. 
\item Solve $\frac{\alpha}{\eta}\{a+\frac{\alpha+1}{\eta+1}\{b+\frac{\alpha+2}{\eta+2}c\}\}-\mu=0$
for $\alpha$, either by solving the cubic, or (better) using Newton-Raphson iteration starting from $\alpha=\eta/2$.
\item find $\beta=\eta-\alpha$ and compute the log-likelihood as usual.
\end{enumerate}
\subsection{Mode}
The mode can be found by differentiating (\ref{eq:cubf}) as for the quadratic distribution.
Then the mode $x_m=ap_m+bp_m^2+cp_m^3$ where
\begin{equation}\frac{\alpha-1}{p_m}-\frac{\beta-1}{1-p_m}-\frac{2b+6cp_m}{a+2bp_m+3cp_m^2}=0.\label{eq:cmode}\end{equation}
One can solve by Newton-Raphson iteration or by solving the resulting cubic equation for $p_m$.
Compared with the beta distribution, this distribution has a different modal structure, unlike the Q-beta distribution, which had the same structure as the beta.
For $c \ne 0$ there is always one mode in $(0,1)$, besides the fact that the pdf will be infinite at zero if $\alpha < 1$ and at unity if $\beta < 1$.
This behaviour, giving a mode superimposed on a U or J-shaped distribution, while interesting, is probably not often wanted.
It was this behaviour, caused by the fact that the Jacobian can be small over a range of $x$, that led to the creation of the `Jacobian-less distribution, described on the next section.

Modal regression could be done by taking parameters $x_m, \eta=\alpha+\beta, \gamma$ and $\delta$. Given the mode $x_m$ and setting $\beta=\eta-\alpha$ one 
finds $p_m$ and then solves (\ref{eq:cmode}) for $\alpha$, which only requires solving a linear equation. Then the likelihood can be computed in terms of $\alpha, \beta, a, b, c$
by setting $\beta=\eta-\alpha$. 
\section{Distributions lacking the Jacobian: SQ-beta and SC-beta distributions}
The form of this distribution, as $Cf_p(p)(a+2bp+3cp^2)$, was derived earlier.

Making $g_x(x)$ integrate to unity gives
\[C^{-1}=a+2b \frac{\alpha}{\alpha+\beta}+3c\frac{\alpha(\alpha+1)}{(\alpha+\beta)(\alpha+\beta+1)}.\]
The properties of the SC-beta distribution only are described; the SQ-beta distribution is of course similar but simpler.
\subsection{Pdf}
The pdf is
\[g_x(x)=\frac{p(x)^{\alpha-1}(1-p(x))^{\beta-1}}{B(\alpha,\beta)(a+2b \frac{\alpha}{\alpha+\beta}+3c\frac{\alpha(\alpha+1)}{(\alpha+\beta)(\alpha+\beta+1)})}\]
where $a, b, c$ are derived from $\gamma, \delta$ as before and $p(x)$ is defined as before.
\subsection{Distribution function}
From integrating the pdf this is
\[G(x)=C\{aI(\alpha,\beta;p(x))+2b\frac{\alpha}{\alpha+\beta} I(\alpha+1,\beta;p(x))+3c\frac{\alpha(\alpha+1)}{(\alpha+\beta)(\alpha+\beta+1)}I(\alpha+2,\beta;p(x))\}.\]
This form requires three evaluations of the incomplete beta function. However, the computation can be made quicker (but messier) using the identity
\[I(\alpha+1,\beta; x)=I(\alpha,\beta;x)-\frac{x^\alpha(1-x)^\beta}{\alpha B(\alpha,\beta)},\]
which is well-known, and can be derived by integrating $J(\alpha+1,\beta)=\int_0^x u^\alpha(1-u)^{\beta-1}\dee u$ by parts, differentiating the $u^\alpha$ term.
We then have
\[G(x)=C\{a+2b\frac{\alpha}{\eta}+3c\frac{\alpha(\alpha+1)}{\eta(\eta+1)}\}I(\alpha,\beta;x)-\frac{Cx^\alpha(1-x)^\beta}{B(\alpha,\beta)}\{\frac{2b}{\eta}+3c(\frac{\alpha+1}{\eta(\eta+1)}+\frac{x}{\eta+1})\},\]
which requires only one evaluation of an incomplete beta function.
\subsection{Random numbers}
The pdf of the parent distribution is the $Beta(\alpha,\beta)$ pdf multiplied by $M(p)=C(a+2bp+3cp^2)$. The rejection method can be used to generate random numbers, by generating 
random numbers $P$ from $Beta(\alpha,\beta)$ and accepting them with probability $M(P)/M_\text{max}$.
The maximum value of $M$ is either at $p=0$ or $p=1$ or a stationary value in $[0,1]$and so is $CM_{\text{max}}$,
where $M_{\text{max}}$ is either $\text{max}(a,a+2b+3c)$ or its maximum with the stationary value at $-b/3c$, if $0 < -b/3c < 1$. Hence the acceptance probability is $\frac{a+2bP+3cP^2}{M_\text{max}}$
Then, given $P$, one forms $X=aP+bP^2+cP^3$.

The efficiency (proportion of generated random numbers retained) is
\[\int_0^1 f_p(p)\frac{(a+2bp+3cp^2)}{M_\text{max}}\dee p=C^{-1}/M_\text{max}.\]
This varies depending on the parameter values, but for the examples it was 33.8\% and 8\%.

This method of generating random numbers works, but a more efficient method would be desirable. 
However, designing a more efficient method would be another research project; there are many ways to proceed.
\subsection{Moments}
The $n$th moment is
\[\text{E}(X^n)=C\int_0^1 f_p(p)(a+2bp+3cp^2)(ap+bp^2+cp^3)^n \dee p.\]
The mean is then
\[\text{E}(X)=\frac{C\alpha}{\alpha+\beta}\{a^2+\frac{\alpha+1}{\alpha+\beta+1}\{3ab+\frac{\alpha+2}{\alpha+\beta+2}\{4ac+2b^2+\frac{\alpha+3}{\alpha+\beta+3}\{5bc+\frac{\alpha+4}{\alpha+\beta+4}(3c^2)\}\}\}\}.\]
\subsection{Mode}
The mode is simply
\[p_m(x)=\frac{\alpha-1}{\alpha+\beta-2}\]
if it exists, or in full $x_m=ap_m+bp_m^2+cp_m^3$. Modal regression would thus be straightforward. The parameters would be $(x_m, \eta,\gamma, \delta)$ and the equation
for $x_m$ would be solved for $\alpha$, after which $\beta=\eta-\alpha$ and the pdf can be computed.
The transformation used gives rise to two simple distributions that generalise the uniform distribution and allow modal or U-shaped distributions. They are mentioned in appendices A and B for completeness and because
they are new. They may find some use in modelling.
\section{Fitting to data}
Two datasets were fitted. The first comprises 252 observations of calculated percentage of body fat plus a variety of other body size measurements, downloaded from statlib
and supplied by Dr. A. Fisher. It is referenced in Penrose {\em et al} (1985).
The second dataset, also from statlib, is a sample of 349 observations of glycosylated hemoglobin (HBA1c) readings reported in DCCT
percentages from diabetic patients. This is referenced in Daramola (2012).

\begin{figure}
\centering
\makebox{\includegraphics{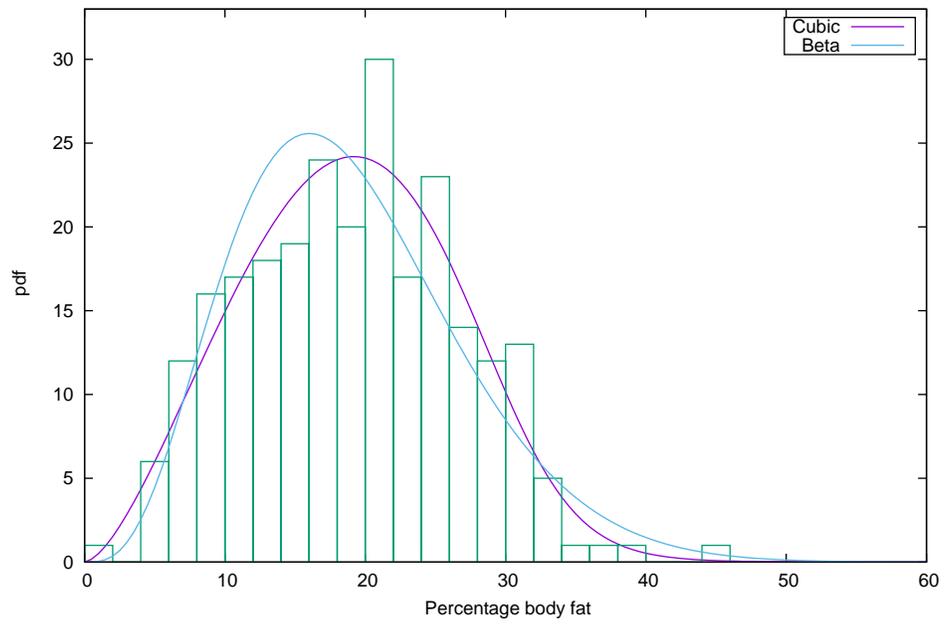}}
\caption{\label{fig1}Beta distribution and C-beta (cubic) beta distributions fitted to the body fat data.
}
\end{figure}

\begin{figure}
\centering
\makebox{\includegraphics{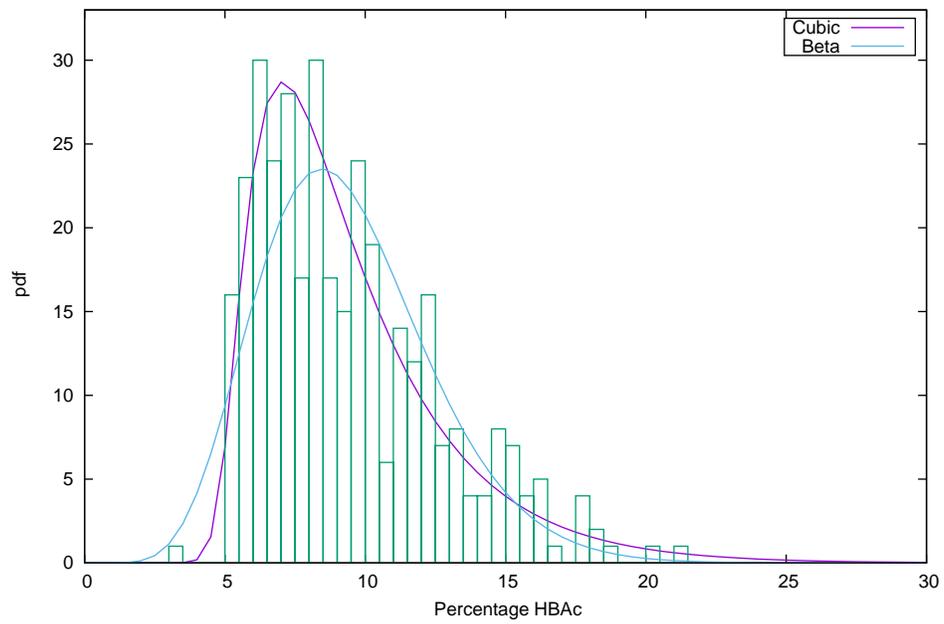}}
\caption{\label{fig2}Beta distribution and C-beta (cubic) beta distribution fitted to the HBA1 data.}
\end{figure}

The results of fitting the beta distribution model, and the Q-beta (quadratic) and C-beta (cubic) models, are shown in table \ref{tab1}.
The Jacobian-less distributions SQ-beta and SC-beta were also fitted.
Figures \ref{fig1} and \ref{fig2} show histograms of the data, with fitted beta and C-beta distributions.
In both cases, the cubic distribution gives a very significant improvement in the log-likelihood. We have in the first case $X^2[2]=10.66, p=0.0048$
and in the second $X^2[2]=33.36, p < 0.001$, showing that the two added parameters significantly improve the fit.

In the first case, the distribution of percentage body fat is almost normal, whereas the beta distribution skews it to the right. The cubic distribution can correct this and give a good fit.
In the second case, the data are more skewed to the right than the beta distribution would allow. The cubic distribution corrects this opposite problem.

The Jacobian-less distributions in fact fit slightly better in both cases, as seen in table \ref{tab1}. The fitted parameters are quite similar.

\section{Conclusions}
The beta distribution has been generalized by allowing quadratic or cubic functions of the beta random variable.
These distributions, especially the cubic, can greatly improve model fit for doubly bounded data.

They are fairly tractable, with moments that are rational functions, which allows a straightforward regression of the mean on covariates,
and are label invariant like the beta distribution. Modes are computable either as solutions of a quadratic/cubic equation or by Newton-Raphson iteration,
so that modal regression is also possible. Distribution functions and random number generation `piggy-back' off that for the parent beta distribution.

An obvious modification is to omit the Jacobian in the transformed  distribution, so that the parent distribution is now a mixture of beta distributions,
where some of the weights can be negative.
The rationale is that for some parameter values, a small Jacobian can introduce an extraneous peak into the distribution.
The modified cubic distribution fitted the examples slightly better than the originals. It has a much simpler expression for the mode,
and is unimodal for $\alpha >1, \beta > 1$
but it has messier expressions for the moments and the distribution function. It would of course also be useful if carrying out modal regression rather than mean regression.

These distributions could be useful in data fitting and as prior distributions. The beta distribution is well-known as the conjugate prior of the binomial distribution,
and a more flexible prior can be useful, \eg for sensitivity analysis.

Obvious future work would be to study the 5-parameter quartic distribution. However, the 4-parameter cubic distribution can already reproduce a wide range of behaviour,
so this is not an urgent task. More efficient generation of random numbers for the SC-beta distribution would be useful.
The method of generalizing the beta distribution proposed here can be applied to any distribution for doubly-bounded data, thus generating
a vast number of possibilities.

\section*{Appendix A: the C-beta$(1,1,\gamma,\delta)$ (2-parameter) distribution}
\begin{figure}
\centering
\makebox{\includegraphics{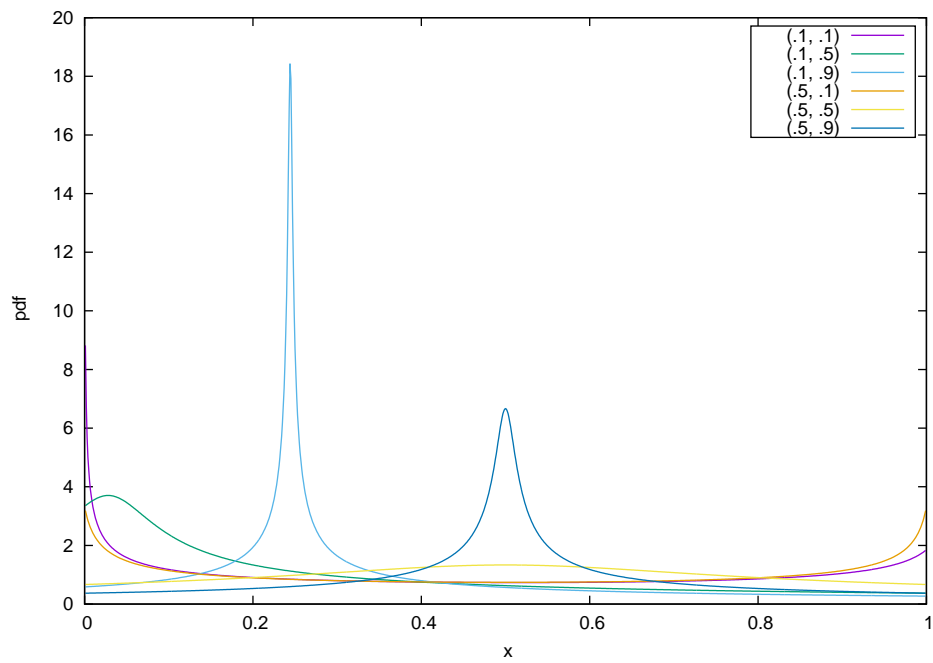}}
\caption{\label{figc}Pdf of the C-beta$(1,1,\gamma,\delta)$  distribution for the six values of $(\gamma,\delta)$ shown in the key.}
\end{figure}

First, the C-beta(1,1) distribution is a special case of the C-beta$(\alpha,\beta)$ distribution discussed earlier. With parameters $\gamma,\delta$, define $a, b, c$ as before.
Then the pdf is $f(x)=1/(a+2bp(x)+3cp(x)^2)$, where as before $p(x)$ solves $x=ap+bp^2+cp^3$. 
Figure \ref{figc} shows the pdf for various values of $\gamma$ and $\delta$.

The distribution function is simply $F(x)=p(x)$.

The moments are
\[\text{E}(X)=a/2+b/3+c/4,\]
\[\text{var}(X)=a^2/12+4b^2/45+9c^2/112+ab/6+bc/6+3ac/20.\]

The mode (which may be an antimode) is at $-b/3c$ if this lies in $(0,1)$. The curvature at the mode is $-6c/J(p)^4$, so if $c > 0$ $(\delta > 1/3)$ the curvature is negative, and it is a mode not an antimode.
For $\delta < 1/2$ the mode may not exist, but for $\delta > 1/2$ it always does.

Random numbers are generated by $X=aU+bU^2+cU^3$ where $U$ is uniform on $[0,1]$.

This distribution with $\delta > 1/2$ can give narrow peaks with a flattish background, and would be suitable as a prior distribution with fat tails.
\section*{A general 2-parameter quadratic distribution}
The Jacobian-less distribution SC-beta$(1,1,\gamma,\delta)$ is simply the uniform distribution. However, its parent, a quadratic distribution, is of interest.
This type of distribution was not considered in the general case, because it did not fit the data as well as the C-beta and SC-beta distributions.
When $\alpha=\beta=1$ however we have a new and potentially useful distribution.

\begin{figure}
\centering
\makebox{\includegraphics{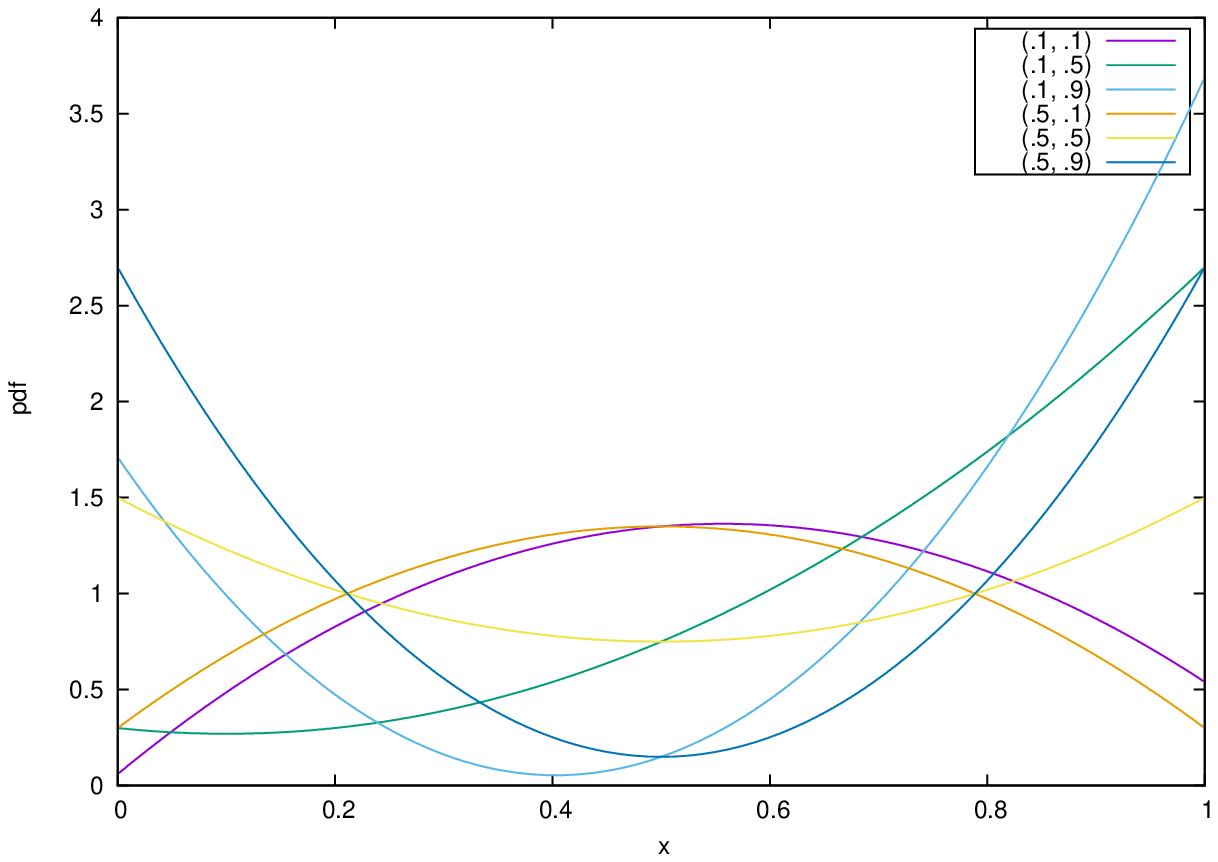}}
\caption{\label{figd}Pdf of the 2-parameter quadratic distribution for the six values of $(\gamma,\delta)$ shown in the key.}
\end{figure}

The distribution has pdf $f(p)=a+2bp+3cp^2$. This is shown in figure \ref{figd} for various values of $\gamma$ and $\delta$.

This gives distribution function $F(p)=ap+bp^2+cp^3$.
Random numbers can be generated in at least two ways. One is to solve
$aP+bP^2+cP^3-U=0$, where $U$ is a uniformly-distributed random number, either analytically or using Newton-Raphson iteration.
The other method is rejection sampling, by generating $U$ and accepting it with probability $f(U)/f_{\text{max}}$, as described for the Jacobian-less distribution.

The moments are
\[\text{E}(P)=a/2+2b/3+3c/4,\]
\[\text{var}(P)=a/3+b/2+3c/5-(a/2+2b/3+3c/4)^2.\]
The moment-generating function can be found from 
\[\text{E}(\exp(tP))=\int_0^1 (a+2bp+3cp^2)\exp(tp)\dee p,\]
as
\[\text{E}(\exp(tP))=a\frac{(\exp(t)-1)}{t}+2b\{\frac{\exp(t)}{t}-\frac{(\exp(t)-1)}{t^2}\}+3c\{\frac{\exp(t)}{t}-2\frac{\exp(t)}{t^2}+2\frac{(\exp(t)-1)}{t^3}\}.\]

The mode is again at $p=-b/3c$ if this lies in $(0,1)$. 

This distribution generalizes the uniform and U-quadratic distributions.

\section*{Figures and tables}
\begin{table}[h]
\begin{tabular}{|l|l|c|c|c|c|c|} \hline
Dataset&Model&$-\ell$&$\alpha$&$\beta$&$\gamma$&$\delta$ \\ \hline
Body fat&Beta&-288.26&4.36&18.67&-&-\\ \hline
Body fat&Q-beta&-288.70&4.27&25.50&.694&-\\ \hline
Body fat& SQ-beta&-288.68&4.28&25.19&.6888&-\\ \hline
Body fat&C-beta&-293.59&2.61&10.95&.354&.637 \\ \hline
Body fat&SC-beta&-293.88&2.63&9.67&.339&.728\\ \hline
HBA1&Beta&-731.48&8.45&81.32&-&-\\ \hline
HBA1&Q-beta&-735.80&15.50&50.51&.1024&-\\ \hline
HBA1&SQ-beta&-735.66&15.23&51.21&.0994&-\\ \hline
HBA1&C-beta&-748.16&14.64&19.56&.057&.641\\\hline
HBA1&SC-beta&-749.35&13.09&19.30&.041&.682\\ \hline
\end{tabular}
\caption{\label{tab1}Results of model fitting to the body fat and HBA1 data. SQ-beta and SC-beta denote the quadratic cubic model omitting the Jacobian.}
\end{table}



\begin{thebibliography}{99}
\bibitem{bayes}Bayes, C. L., Baz\'{a}n, J. L. and Garcia, C. (2012). A new robust regression model for proportions, Bayesian Analysis, {\bf 7}, 841-866.
\bibitem{daramola}Daramola O.F. (2012).  Assessing the validity of random blood glucose testing
for monitoring glycemic control and predicting HbA1c values in type 2 
diabetics at Karl Bremer hospital. Masters Thesis (Family Medicine and
 Primary Care), Stellenbosch University: Stellenbosch, South Africa.
\bibitem{gomez}G\'{o}mez-D\'{e}niz, E., Sordo, M. A. and Calder\'{i}n-Ojeda, E. (2014), The log-Lindley distribution as an alternative to the beta regression model with applications in insurance, Insurance: mathematics and Economics, 
{\bf 54}, 49-57.
\bibitem{johnson}Johnson, N. L., Kotz, S. and Balakrishnan, N. (1995). Continuous Univariate Distributions, Wiley, New York.
\bibitem{beyond}Kotz, S. and van Dorp, J. R. (2004). Beyond Beta: Other Continuous Families of Distributions with Bounded Support and Applications,
World Scientific, Singapore.
\bibitem{kum}Kumaraswamy, P. (1980). A generalized probability density function for double-bounded random processes, Journal of Hydrology {\bf 46} (1-2), 79–88.
\bibitem{mielke}Mielke, P. W. Jr. (1975). Convenient beta distribution likelihood techniques for describing and comparing meteorological data, Journal of Applied Meterology, {\bf 14}, 985-990.
\bibitem{nada}Nadarajah, S. and Kotz, S. (2006). Beta trigonometric distributions, Portuguese Economic Journal, {\bf 5}, 207-224
\bibitem{pen}Penrose, K., Nelson, A., and Fisher, A. (1985). Generalized Body
Composition Prediction Equation for Men Using Simple Measurement
Techniques" (abstract), Medicine and Science in Sports and Exercise, {\bf 17} (2), 189-189.
\bibitem{pham}Pham-Gia, T. and Duong, Q. P. (1989). The generalized beta and F-distributions in statistical modelling, Mathematical computer modelling, {\bf 12}, 1613-1625.
\bibitem{press}Press, W. H., Teukolsky, S. A., Vetterling, W. T. and Flannery, B. P. (2007). Numerical Recipes: the art of scientific computation, 3rd ed., Cambridge University Press, Cambridge
\bibitem{senn}Senn, S. (1996). Relation between treatment benefit and underlying risk in meta-analysis - Standard of 'label invariance' should not abandoned
British Medical Journal, {\bf 313}, 1550-1550.
\bibitem{stewart}Stewart, C. (2013). Zero-inflated beta distribution for modeling the proportions in quantitative fatty acid signature analysis, Journal of Applied Statistics, {\bf 40}, 985-992.
\bibitem{vandorp}Van Dorp, J R. and Kotz, S. (2002). The standard two-sided power distribution and its properties: with applications to financial engineering, American Statistician, 
{\bf 56}, 90-99.
\end{thebibliography}
\end{document}